%% file: main.tex
\documentclass[9pt, conference]{IEEEtran}
\IEEEoverridecommandlockouts
\usepackage{cite}
\usepackage{amsmath,amssymb,amsfonts}
\usepackage{algorithmic}
\usepackage{graphicx}
\usepackage{svg}
\usepackage{textcomp}
\usepackage{xcolor}
\usepackage{url}
\usepackage{multirow}
\usepackage[caption=false,font=footnotesize]{subfig}
\usepackage{tikz}
\usepackage{array}
\usepackage{booktabs}
\usepackage{colortbl}
\usepackage{siunitx}
\usepackage{listings}
\usepackage{enumitem}

\definecolor{llmwin}{RGB}{198,234,208}
\lstdefinestyle{ieeemd}{
  basicstyle=\ttfamily\scriptsize,
  breaklines=true,
  breakatwhitespace=false,
  columns=fullflexible,
  keepspaces=true,
  frame=single
}

\def\BibTeX{{\rm B\kern-.05em{\sc i\kern-.025em b}\kern-.08em
T\kern-.1667em\lower.7ex\hbox{E}\kern-.125emX}}

\begin{document}
\emergencystretch=1em
\hbadness=10000
\vbadness=10000
\hfuzz=1pt

\title{CHICO-Agent: An LLM Agent for the Cross-layer Optimization of 2.5D and 3D Chiplet-based Systems}

\author{Qihang Wu, Aman Arora, and Vidya A. Chhabria \\ Arizona State University}


\maketitle

\input{sec/1-abstract.tex}

\input{sec/3-intro.tex}
\input{sec/4-background.tex}
\input{sec/5-overview.tex}

\input{sec/6-chico-agent.tex}

\input{sec/7-experiment.tex}

\input{sec/8-results.tex}

\input{sec/10-conclusion.tex}
\newpage

\bibliographystyle{bib/ieeetr2}
\bibliography{bib/bibfile}

\end{document}

%% file: sec/1-abstract.tex
\begin{abstract}
The rapid growth of large language models (LLMs) and AI workloads has pushed monolithic silicon to its reticle and economic limits, accelerating the adoption of 2.5D/3D chiplet systems. However, these systems increase design complexity by requiring co-design across multiple levels of the computing stack, including application, architecture, chip, and package. The resulting design space is highly combinatorial, with trade-offs among latency, energy, area, and cost. To address this challenge, we propose CHICO-Agent, an LLM-driven optimization framework for 2.5D/3D chiplet-based systems. CHICO-Agent maintains a persistent knowledge base to capture parameter–outcome trends and coordinates exploration through an admin–field multi-agent workflow. Compared with a simulated-annealing baseline, CHICO-Agent finds lower-cost configurations and provides an interpretable audit trail for designers.


\end{abstract}

%% file: sec/3-intro.tex
\section{Introduction}
\begin{figure}
  \centering
  \includegraphics[width=0.8\linewidth]{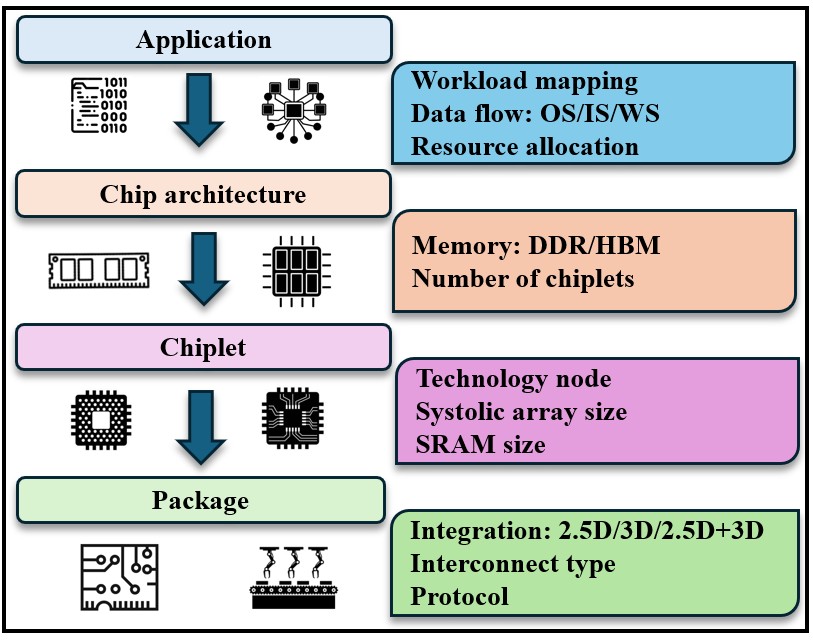}
  \vspace{-3mm}
  \caption{Computing stack highlighting the searchable design space spanning different levels, for 2.5D and 3D chiplet-based systems.}
  \vspace{-5mm}
  \label{fig:design_space_rep}
\end{figure}

\noindent
\textbf{Motivation.} 
The rapid growth of artificial intelligence (AI) workloads has pushed monolithic silicon to its physical and economic limits, driving the adoption of heterogeneous integration (HI) using 2.5D and 3D chiplet-based architectures. In these systems, designs are partitioned into smaller dies interconnected through advanced packaging to improve yield, reduce costs, and enable chiplet reuse. However, HI systems require \emph{cross-layer co-design} across application, architecture, chiplet, and packaging levels, as illustrated in Fig.~\ref{fig:design_space_rep}. Decisions at one layer influence others; for example, workload partitioning affects communication patterns, which impact interconnect topology, packaging, and metrics such as latency, energy, area, and cost. The many variables across these layers create a combinatorial design space, making exhaustive exploration infeasible and necessitating automated design space exploration (DSE) and optimization.



\noindent
\textbf{Limitations of Existing Exploration Methods.} Prior work has developed modeling and simulation frameworks to estimate system-level metrics, including performance, power, area, and cost (PPAC), for chiplet-based architectures. \textit{HISIM}~\cite{hisim} provides analytical models for evaluating performance trade-offs in 2.5D and 3D systems, while \textit{CATCH}~\cite{catch} focuses on cost modeling. These frameworks enable evaluation of candidate architectures. 
However, relatively little work has addressed optimization and exploration of chiplet design spaces. As architectural, chiplet, and packaging parameters grow, manual exploration and parameter sweeps become infeasible. 

A few works explore automated DSE for chiplet systems. \textit{ChipletGym}~\cite{chiplet-gym} formulates exploration as a reinforcement learning problem, while \textit{CarbonPATH}~\cite{carbon-path} uses simulated annealing (SA). However, these approaches rely on stochastic search or metaheuristics that explore the space through random perturbations and iterative evaluation~\cite{carbon-path, chiplet-gym}. As a result, they can become trapped in local minima and often require careful hyperparameter tuning to obtain high-quality solutions. Moreover, many black-box search methods used for chiplet DSE do not maintain an explicit, human-interpretable knowledge, which limits how directly designers can reuse discovered cross-layer patterns. This limits the designer and/or optimizer to exploit relationships across architecture, chiplet, and packaging parameters.

\noindent
\textbf{LLM-guided Reasoning for Design Space Exploration.}
Recent advances in large language models (LLMs) have demonstrated strong capabilities in reasoning, planning, and iterative problem solving~\cite{ReAct}. These models can synthesize structured knowledge, identify relationships among variables, and refine solutions based on historical observations, creating an opportunity to rethink traditional hardware design optimization and design space exploration strategies. Instead of relying solely on stochastic perturbations, LLMs can guide the search by reasoning over previously evaluated configurations and identifying promising regions of the design space.

LLMs have also been explored for DSE in chip design. \textit{MAHL}~\cite{MAHL} uses multi-agent LLMs to generate hierarchical chiplet RTL designs with adaptive debugging, targeting synthesis and PPA optimization. \textit{3D-CIMlet}~\cite{3D-CIMlet} proposes a co-design framework for 3D compute-in-memory chiplets for edge LLM inference. \textit{ChatNeuroSim}~\cite{ChatNeuroSim} employs LLM agents to automate compute-in-memory simulator workflows with design space pruning. However, these works address different design problems. Our work focuses on cross-layer system configuration, identifying chiplet counts and types, memory technologies, workload mapping strategies, packaging architectures, and interconnect choices to satisfy PPAC constraints for a given workload.

\noindent
\textbf{Our Contribution.}
CHICO-Agent uses LLM reasoning to perform cross-layer DSE across the compute stack by iteratively analyzing historical PPAC evaluations to identify parameter–metric relationships and propose optimized configurations. The contributions are:

\begin{itemize}

\item We develop \textbf{CHICO-Agent}, an LLM-guided framework for cross-layer optimization of chiplet architectures that reasons over previously evaluated designs to guide exploration.


\item CHICO-Agent and its LLM reasoning can identify higher-quality system configurations than metaheuristic baselines.

\item CHICO-Agent significantly reduces the hyperparameter tuning burden inherent to traditional metaheuristics.

\item We show that the reasoning traces generated by the LLM provide interpretable rationales linking design parameter changes to PPAC outcomes, offering actionable insights for designers.

\end{itemize}

\noindent CHICO-Agent is available on an anonymized GitHub repository~\cite{CHICO-Agent}.

%% file: sec/4-background.tex
\section{Preliminaries}
\label{sec:background}

\noindent
\textbf{Analytical Modeling of Chiplet-Based Systems}
\label{subsec:analytical_modeling}
Designing chiplet-based systems requires evaluating a large number of candidate system configurations that span multiple design dimensions. Accurately evaluating each system configuration using detailed simulation or full-system modeling can be computationally expensive, making such approaches impractical for large-scale DSE. To address this challenge, prior work has developed analytical models that estimate system-level metrics as functions of architectural and packaging parameters~\cite{chiplet-gym, carbon-path, hisim}. Analytical models provide fast approximations of PPAC metrics such as system performance, power, area, and costs by developing models across the compute stack. 
In our work, we leverage existing analytical models from~\cite{carbon-path} to model PPAC, and we focus on an optimization framework that combines these models with LLMs to search the design space.


\noindent
\textbf{Optimization and DSE for Chiplet-based Systems.}
The DSE problem is formulated as an optimization problem, with the goal of minimizing a cost function given the \emph{target workload} and \emph{application profile}. The target workload specifies the computation to be accelerated (e.g., a GEMM operator defined by its tensor dimensions and batch size); it determines the volume of computation, data movement, and inter-chiplet communication, and therefore directly influences all PPAC metrics. The application profile assigns weighting coefficients to the PPAC objectives, reflecting deployment-specific priorities---for instance, a wearable application emphasizes energy and area, whereas an automotive application prioritizes latency and manufacturing cost. The cost function integrates multiple design objectives into a single scalar metric, enabling the optimizer to compare alternative solutions quantitatively. In CHICO-Agent, the cost function is:

\vspace{-2mm}
{\footnotesize
    \begin{equation}
    \begin{split}
        \text{System Cost}(\mathbf{x}) ={}& \alpha \, E_\text{system}(\mathbf{x})
            + \beta \, A_\text{system}(\mathbf{x}) \\
            &+ \gamma \, L_\text{system}(\mathbf{x})
            + \theta \, C_\text{system}(\mathbf{x})
    \end{split}
    \label{eq:system_cost}
    \end{equation}
}

\noindent where $\mathbf{x}$ is a system configuration, and $\alpha$, $\beta$, $\gamma$, and $\theta$ are application-specific weighting coefficients for energy, area, latency, and manufacturing cost, respectively. Formally, the goal is to find a system configuration $\mathbf{x}$ that minimizes the system cost for the specific workload and optimization profile:

\vspace{-3mm}
\begin{equation}
    \mathbf{x}^{*} = \arg\min_{\mathbf{x} \in \mathcal{X}} \; \text{System Cost}(\mathbf{x})
    \vspace{-1mm}
    \label{eq:opt_problem}
\end{equation}

\noindent
where $\mathcal{X}$ denotes the set of feasible configurations. Each of the terms in the cost function are unit-converted and normalized to their median value of 10,000 different system configurations of $\mathbf{x}$~\cite{carbon-path}. The configuration vector $\mathbf{x}$ is a selection of design parameters spanning four levels of the compute stack as defined in Fig.~\ref{fig:design_space_rep} and described below~\cite{carbon-path}:

\begin{itemize}[nosep,leftmargin=*]
    \item \textit{Application \& Mapping:} the \emph{dataflow strategy} (output-stationary (OS), weight-stationary (WS), or input-stationary (IS)) determines which tensor operand is maximally reused in the on-chip buffer; the \emph{workload assignment order} (ascending or descending) controls whether GEMM tiles are allocated starting from the smallest or largest chiplet, directly affecting load balance in heterogeneous configurations where chiplets differ in compute throughput; \emph{Split-K partitioning} (on/off) decides whether the reduction dimension~$K$ is split across chiplets, introducing inter-chiplet communication for partial-sum aggregation in exchange for higher parallelism; and \emph{chiplet data sharing} (on/off) enables or disables the broadcast-based reuse of input data across chiplets, converting redundant DRAM accesses into cheaper die-to-die transfers. We use the shorthand \textit{O\nobreakdash-D\nobreakdash-K} to denote a workload mapping, where \textit{O} is the assigning order (0\,=\,descending, 1\,=\,ascending), \textit{D} the dataflow (OS\,=\,output-stationary, WS\,=\,weight-stationary, IS\,=\,input-stationary), and \textit{K} indicates split-K (1\,=\,enabled, 0\,=\,disabled); e.g., \textit{1\nobreakdash-OS\nobreakdash-0}.

    \item \textit{Architecture:} the \emph{number of chiplets} ($1$--$6$) and the \emph{system memory type} (DDR4, DDR5, HBM2, HBM3) determine bandwidth and energy per access, directly influencing data movement cost.

    \item \textit{Chiplet:} the \emph{technology node} (7, 10, 14\,nm) governs transistor density, leakage power, and cost per unit area; the \emph{systolic array size} ($64{\times}64$ to $192{\times}192$) defines the throughput per chiplet; and the \emph{on-chip SRAM buffer capacity} controls the volume of data that can be reused locally before accessing DRAM memory. We denote a chiplet configuration as \textit{A\nobreakdash-T\nobreakdash-S}, where \textit{A} is the systolic array dimension, \textit{T} is the technology node (nm), and \textit{S} is the SRAM buffer capacity (KB); for example, \textit{64\nobreakdash-7\nobreakdash-512} denotes a 64$\times$64 array at 7\,nm with 512\,KB SRAM.

    \item \textit{Package:} the \emph{integration style} (2D, 2.5D, 3D, 2.5D+3D) determines whether chiplets are arranged side-by-side, stacked vertically, or in a hybrid configuration; the \emph{interconnect type} (e.g., RDL, EMIB, TSV, hybrid bond) sets the physical die-to-die link technology, governing bandwidth density and energy per bit; the \emph{communication protocol} (e.g., UCIe, AIB, BoW) defines the signaling standard and its associated data rate; and the \emph{die-to-die topology} specifies how chiplets are logically connected (e.g., ring, mesh, star), affecting hop count and aggregate bandwidth. We summarize packaging as \textit{I\nobreakdash-P\nobreakdash-M}, where \textit{I} is the integration style, \textit{P} the interconnect, and \textit{M} the memory type; e.g., \textit{2.5D\nobreakdash-RDL\nobreakdash-DDR5}. We abbreviate active and passive interposers as \textit{Acti} and \textit{Pass}, microbump and hybrid bond as \textit{\textmu B} and \textit{HB}, and die-to-die protocols as Universal Chiplet Interconnect Express (UCIe) 3D (UC3), UCIe Standard (UCS), and UCIe Advanced (UCA).
    
\end{itemize}

\noindent
The analytical models from~\cite{carbon-path} account for these parameters and their cross-layer interactions when computing each PPAC for a given $\mathbf{x}$. Our goal in CHICO-Agent is to use these models and solve Eq.~\eqref{eq:opt_problem}.

%% file: sec/5-overview.tex
\section{CHICO-Agent Overview}
\begin{figure}[t]
    \centering
    \includegraphics[width=0.8\columnwidth, keepaspectratio]{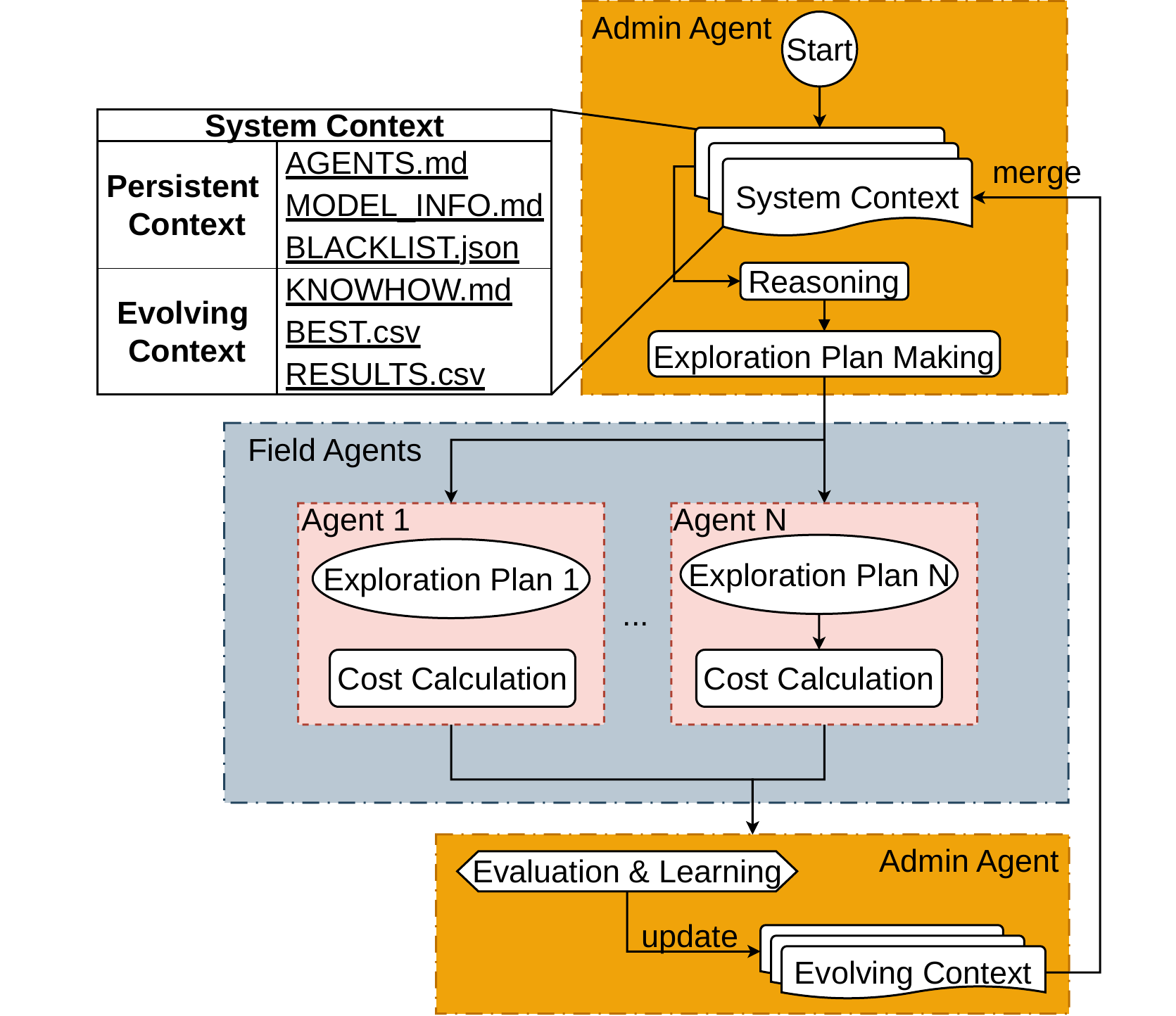}
    \vspace{-3mm}
    \caption{Hierarchical multi-agent iterative framework of CHICO-Agent with admin agents and field agents. $N$ (default\,=\,100) sets the number of parallel field agents spawned per iteration, and \emph{max iterations} governs how many complete iterations the framework executes.}
    \vspace{-2mm}
    \label{fig:admin_field}
\end{figure}

\noindent
CHICO-Agent is a hierarchical multi-agent framework inspired by~\cite{self-driving_codebase} that solves the optimization problem defined earlier and is illustrated in Fig.~\ref{fig:admin_field}. The framework operates iteratively through an \textit{admin agent} and a set of \textit{field agents}. In each iteration, the admin agent reasons over the current \textit{system context} and generates $N$ exploration plans. $N$ field agents execute these plans in parallel and return their results, after which the admin agent consolidates the findings and updates the context for the next iteration. This feedback loop enables the system to accumulate insights over \textit{max iterations}, progressively narrowing the search toward high-quality design points. Rather than relying on stochastic metaheuristics, we treat the optimization task as a reasoning problem, where an LLM interprets PPAC metrics and adapts its search strategy based on previously observed outcomes.

The framework maintains two types of context. The \textit{persistent context} provides static domain knowledge, including \texttt{AGENTS.md} (agent roles and workflow), \texttt{MODEL\_INFO.md} (PPAC modeling information), and \texttt{BLACKLIST.json} (invalid architectural combinations). The \textit{evolving context} grows with each iteration and captures exploration history through \texttt{KNOWHOW.md}, \texttt{BEST.csv}, and \texttt{RESULTS.csv}. Together, these form the \textit{system context} used by the admin agent at the start of each iteration as shown in Fig.~\ref{fig:admin_field}.

%% file: sec/6-chico-agent.tex
\section{CHICO-Agent Detailed Architecture}\label{sec:elaboration}
\noindent
While methods such as simulated annealing~\cite{carbon-path}, particle swarm optimization~\cite{particle-swarm-opt}, and reinforcement learning~\cite{chiplet-gym} traverse high-dimensional design spaces without semantic understanding of the underlying hardware trade-offs, our framework introduces a \emph{context-aware exploration} paradigm using the reasoning capabilities of LLMs. Furthermore, metaheuristics require calibration of hyperparameters such as initial temperature, cooling schedule, and mutation strategies—parameters that are problem-specific and often require extensive manual tuning. In contrast, CHICO-Agent replaces these with a reasoning-driven loop that requires minimal tuning. CHICO-Agent exposes only three parameters: (1) the \emph{reasoning effort}, controls the model’s reasoning budget during planning; (2) the \emph{number of iterations}, which determines how many admin–field exploration iterations are executed; and (3) the \emph{number of field agents} $N$, which governs the degree of parallel exploration per iteration.

\begin{figure}[t]
  \centering
  \includegraphics[width=\linewidth, keepaspectratio]{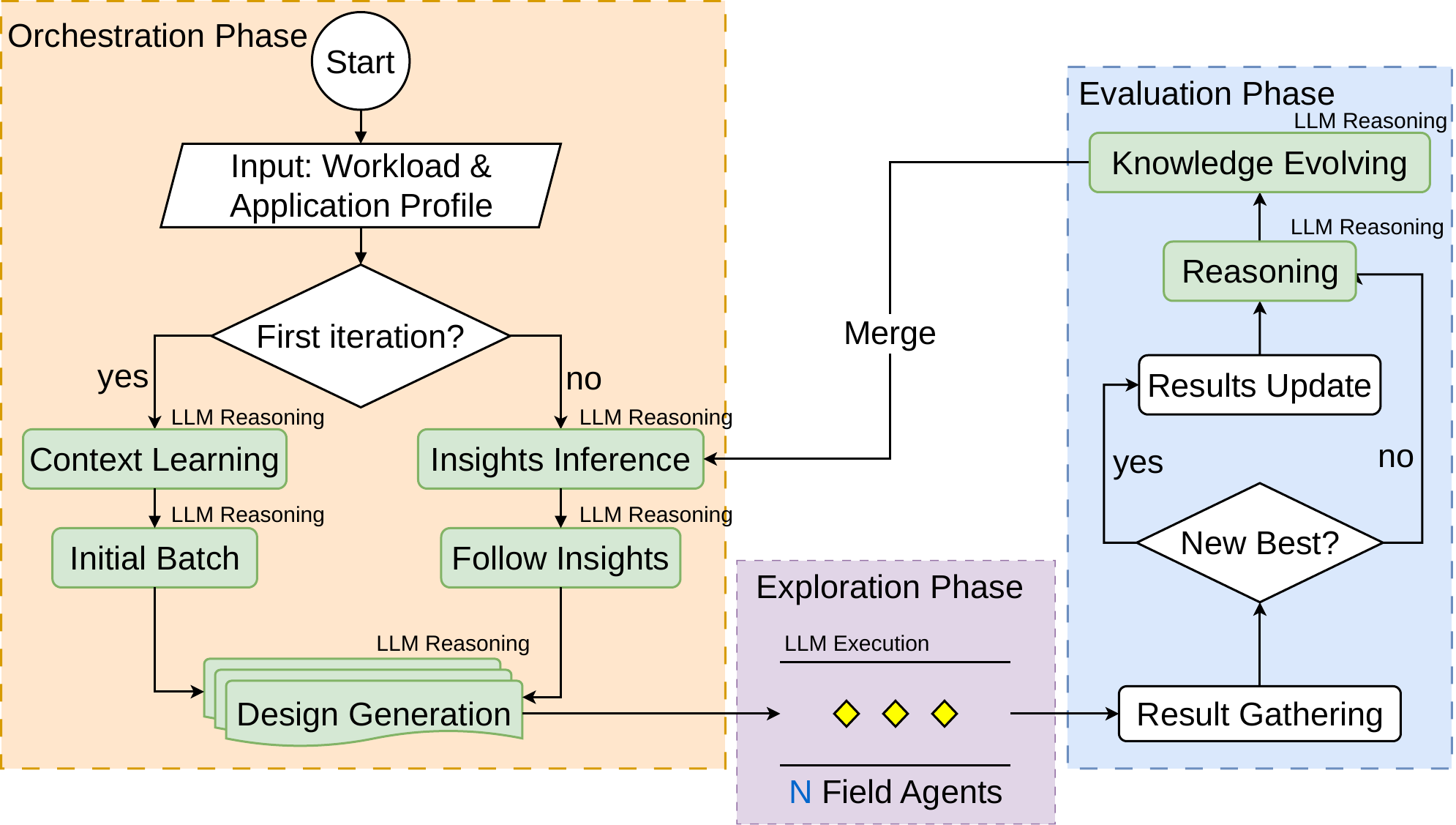}
  \vspace{-5mm}
  \caption{The admin agent iteration loop with three phases. 
  Green rectangles denote stages where the LLM performs active reasoning, whose depth is controlled by the \emph{reasoning effort} hyperparameter. }
  \vspace{-5mm}
  \label{fig:iteration_loop}
\end{figure}

We define an \emph{iteration} as one cycle of the CHICO-Agent optimization loop, as illustrated in Fig.~\ref{fig:iteration_loop}. Each iteration comprises three phases: (1)~an \emph{orchestration phase} where the admin agent ingests the system context, performs reasoning, and generates $N$ exploration plans; (2)~an \emph{exploration phase} where $N$ field agents evaluate plans; and (3)~an \emph{evaluation phase} in which the admin agent gathers field agent results, reasons over them, and updates the evolving context.


\vspace{-1mm}
\subsection{Orchestration Phase}
\label{sec:orchestration}
\noindent
At the start of each iteration, the admin agent ingests the system context and performs structured reasoning to generate $N$ exploration plans targeting distinct regions of the HI design space. The system context includes two components: the \emph{persistent context}, which provides static domain knowledge and behavioral constraints, and the \emph{evolving context}, which accumulates empirical evidence from prior iterations. The process begins with two inputs defined in Section~\ref{sec:background}: the target workload specification (operator type, tensor dimensions, and batch size) and the application profile (PPAC weighting coefficients $\alpha$, $\beta$, $\gamma$, $\theta$), which define the optimization instance in Eq.~\eqref{eq:opt_problem}.

The admin agent then branches based on whether the current iteration is the first. In the initial iteration, no prior exploration history exists, so the admin agent generates an \textit{initial batch} of exploration plans by reasoning over the workload, application profile, and persistent context (PPAC models, blacklist constraints, and design reference tables; Fig.~\ref{fig:admin_field}), providing broad coverage of the HI design space. In subsequent iterations, the admin agent additionally incorporates the evolving context and performs \textit{insights inference} to identify patterns from prior iterations. It then follows these insights to bias plan generation toward promising regions of the design space, transitioning the search from broad exploration to informed exploitation.

\noindent
\underline{\textbf{Persistent Context.}}
\label{sec:persistent}
The persistent context functions as the foundational operating procedure and inference knowledge base for the agent. It defines the objectives and operational constraints for rigorous chiplet DSE. This context provides the admin agent with the domain grounding necessary to orchestrate the optimization and numerically interpret evaluation results via inference and chain of thought (CoT) reasoning~\cite{cot}. The persistent context is maintained as three structured parts: \texttt{AGENTS.md}, \texttt{MODEL\_INFO.md}, and \texttt{BLACKLIST.json}. We describe these three parts below.

\noindent
\underline{\textit{(a) Admin and Field Agentic Workflow.}}
\texttt{AGENTS.md} encodes agent roles and behavior guidelines. It serves as the complete operating procedure for the agent. It specifies: (i) the optimization objective, namely minimizing the weighted PPAC cost for a given GEMM workload and application profile; (ii) hard constraints that restrict the agent's writable scope, fix the cost evaluation interface, prohibit source code modifications, and mandate that all design decisions be evidence based and grounded in the PPAC parameter relationship documentation; (iii) design reference tables listing valid SRAM to array size mappings, protocol--package compatibility rules, and chiplet area/power computation formulas; (iv) the blacklist validation requirement, directing the agent to check every proposed configuration against \texttt{BLACKLIST.json} before evaluation; (v) a structured iteration strategy that prescribes broad exploration during the initial iteration (e.g., sweeping chiplet counts) and evidence driven refinement in subsequent iterations using cumulative findings from \texttt{KNOWHOW.md}; and (vi) a step by step evaluation cycle covering architecture generation, cost evaluation, \texttt{BEST.csv} updates, and the required \texttt{KNOWHOW.md} entry template.


\noindent
\underline{\textit{(b) PPAC Analytical Models.}}
\label{sec:modeling}
CHICO-Agent is provided with a summary of the PPAC models used to evaluate cost in Eq.~\eqref{eq:system_cost} from~\cite{carbon-path}. For brevity, we provide only a high-level overview; detailed parameter definitions, equations, and sources are available in~\cite{carbon-path} and are used by the model as part of its persistent context. 

\noindent
\textit{(i) System Latency Model.} System latency comprises compute, DRAM access, and die-to-die (D2D) communication components. Compute and DRAM read latencies are evaluated per chiplet using a cycle-accurate systolic array simulator~\cite{scale-sim} under the selected dataflow mapping, with system latency determined by the slowest chiplet. D2D latency is modeled as total interchiplet data volume divided by effective link bandwidth, determined by signaling rate, number of bumps, and protocol efficiency. Bump availability depends on integration style: in 3D, bumps span the chiplet area, while in 2.5D they are limited to chiplet edges. DRAM write latency is similarly bounded by the slowest chiplet.

\noindent
\textit{(ii) Energy Model.} Total system energy includes compute energy and D2D communication energy. Compute energy aggregates per chiplet costs of DRAM reads, MAC operations, SRAM accesses, and DRAM writes for the workload. D2D communication energy sums the per-bit transfer energy multiplied by data volume across interchiplet links.

\noindent
\textit{(iii) Area Model.} The system area footprint is estimated via bipartition-based floorplanning of chiplet placements. Monolithic and 3D systems are bounded by base die area, while 2.5D and hybrid 2.5D+3D systems additionally account for interposer or package footprint, including whitespace overhead in chiplet floorplans.

\noindent
\textit{(iv) Manufacturing Cost Model.} This aggregates chiplet fabrication costs, package costs,  and memory costs, accounting for yield.

\noindent
\underline{\textit{(c) Blacklist of Infeasible System Configurations.}}
\label{sec:blacklist}
As formulated in Eq.~\ref{eq:opt_problem}, the optimization searches over the feasible set $\mathcal{X}$. Not all parameter combinations produce physically realizable systems, so invalid configurations must be excluded from $\mathcal{X}$. Examples include mismatched protocol assignments (e.g., using a 3D-only protocol such as UCIe-3D in a 2.5D system), unstable 3D stacks (e.g., stacking a larger die on a smaller one), and incorrect HI classifications (e.g., assigning a 2.5D+3D integration type to a two-chiplet system). To enforce these constraints, \texttt{BLACKLIST.json} is included in the persistent context and enumerates illegal combinations.


\noindent
\underline{\textbf{Evolving Context.}}
\label{sec:evolving}
\noindent
As shown in Fig.~\ref{fig:admin_field}, the evolving context grows across iterations, using \texttt{KNOWHOW.md} as an append-only record of exploration history, \texttt{BEST.csv} as a ranked list of the best (least cost) architectures, and \texttt{RESULTS.csv} as a log of all evaluated configurations with their PPAC metrics. In each iteration, field agents append new learnings to \texttt{KNOWHOW.md}, enabling the admin agent to infer cross-parameter relationships, such as how chiplet count interacts with interconnect protocol selection to affect latency and cost across evaluation batches. The record tracks numerical deltas showing how parameter changes (e.g., SRAM capacity or interconnect protocol) impact PPAC metrics. Guided by these results, the admin agent adjusts its exploration toward lower-cost regions of the design space in subsequent iterations. The top-performing architectures in \texttt{BEST.csv} serve as knowledge for iterative refinement. From the second iteration, we combine the evolving context with the persistent context to form the complete system context (Fig.~\ref{fig:admin_field}).

\noindent
\underline{\textbf{Reasoning and Plan Generation.}}
\label{sec:admin_reasoning}
\noindent
Using the system context, the admin agent performs structured reasoning to generate the next batch of $N$ exploration plans. The agent analyzes \texttt{RESULTS.csv} to identify the current best architecture based on minimum cost and validates its reasoning against the persistent context. By comparing the current batch with the exploration history in \texttt{KNOWHOW.md}, the LLM derives insights from previously evaluated designs. Each reasoning cycle concludes with a next iteration plan, where the agent justifies its strategy using empirical evidence, such as shifting focus from chiplet count to interconnect protocol selection. During plan generation, the agent ensures that no blacklisted system configurations are selected. The resulting $N$ plans are passed to the \textit{Design Generation} stage, which produces HI architectural systems for dispatch to field agents in the exploration phase.


\vspace{-1mm}
\subsection{Exploration Phase}
\label{sec:field}
\noindent
As shown in Fig.~\ref{fig:admin_field}, the exploration phase dispatches the admin agent's $N$ exploration plans to independently spawned field agents, scaling exploration throughput with available compute resources and enabling concurrent evaluation of disjoint regions of the HI design space. Each field agent performs autonomous exploration within its assigned subspace, running simulations and querying performance models to evaluate candidate designs. It records the minimum cost and the corresponding configuration, writes the PPAC results to \texttt{RESULTS.csv}, and produces a natural-language summary of its findings in \texttt{KNOWHOW.md}.


\vspace{-1mm}
\subsection{Evaluation Phase}
\label{sec:eval_learn}
\noindent
During the evaluation phase, the parallel field agent branches converge and their results are gathered by the admin agent, which performs two functions. First, in the \textit{Reasoning} stage, the admin agent analyzes the field agents' findings in the context of prior knowledge, validating its reasoning against PPAC parameter relationship documentation. Second, in the \textit{Knowledge Evolving} stage, the admin agent determines whether a new best-performing design has been discovered. CHICO-Agent follows a ``go with the winners'' strategy~\cite{go-with-the-winners}: if a new best is found, the \textit{Results Update} stage commits the improved configuration to \texttt{BEST.csv}. The workflow then proceeds to the \textit{Merge} stage, where the current iteration's findings are consolidated with existing knowledge, updating \texttt{KNOWHOW.md} and \texttt{RESULTS.csv}. The admin agent then merges all findings into the evolving context, completing the evaluation phase. If the maximum number of iterations has not been reached, the updated system context (Fig.~\ref{fig:admin_field}) feeds into the next iteration, returning to the orchestration phase and closing the loop shown in Fig.~\ref{fig:iteration_loop}.

%% file: sec/7-experiment.tex
\section{Experiment Setup}
\label{sec:experiment}
\noindent

\noindent
{\bf Metrics.} Optimization inherently involves a tradeoff between solution quality and runtime: while local optima can be reached quickly, discovering better solutions requires increased exploration. Therefore, we evaluate CHICO-Agent and the SA baseline in terms of  (1)~\emph{best system cost} achieved, and (2)~\emph{wall-clock time} required to reach it. 


\noindent
{\bf Workload and Optimization Profiles.}
We evaluate six representative GEMM workloads spanning diverse compute and memory characteristics (Table~\ref{tab:gemm_dims}). To model different deployment scenarios, we use application-specific cost weights defined in Table~\ref{tab:optimization_profile_table}.

\vspace{-2mm}
\begin{table}[htbp]
  \centering
  \caption{GEMM workloads for different benchmarks.}
  \resizebox{\linewidth}{!}{%
    \setlength{\extrarowheight}{2pt}%
    \begin{tabular}{!{\vrule width 2pt}c|l!{\vrule width 2pt}c|c|c!{\vrule width 2pt}}
      \noalign{\hrule height 2pt}
      \textbf{WL} & \textbf{Benchmark} & \textbf{M (Batch)} & \textbf{K (Input)} & \textbf{N (Output)} \\
      \noalign{\hrule height 2pt}
      1 & GPT-2 - MLP (feed-forward)      & 512   & 768   & 3072  \\ \hline
      2 & ViT - MLP (batch=32)            & 6304  & 768   & 3072  \\ \hline
      3 & ViT - MLP (batch=1)             & 197   & 768   & 3072  \\ \hline
      4 & ResNet-50 - FC (classifier)     & 128   & 2048  & 1000  \\ \hline
      5 & VGG-16 - FC (classifier)        & 64    & 4096  & 4096  \\ \hline
      6 & MobileNetV2 - Bottleneck        & 1316  & 24    & 144   \\
      \noalign{\hrule height 2pt}
    \end{tabular}%
  }
  \label{tab:gemm_dims}
   \vspace{-3mm}
\end{table}

\vspace{-2mm}
\begin{table}[htbp]
  \centering
  \caption{Cost function weights in Eq.~\eqref{eq:system_cost} for different profiles.}
  \small
  \resizebox{0.65\linewidth}{!}{%
    \setlength{\extrarowheight}{1pt}%
    \begin{tabular}{!{\vrule width 2pt}c!{\vrule width 2pt}c|c|c|c!{\vrule width 2pt}}
    \noalign{\hrule height 2pt} 
    \textbf{Application} & $\boldsymbol{\alpha}$ & $\boldsymbol{\beta}$& $\boldsymbol{\gamma}$ & $\boldsymbol{\theta}$  \\
    \noalign{\hrule height 2pt} 
    Balance       & 1    &  1  & 1   & 1   \\ \hline
    Mobile        & 0.8  & 0.2 & 0.1 & 0.1 \\ \hline
    Automotive    & 0.1  & 0.1 & 0.7 & 0.7 \\ \hline
    Wearables     & 0.6  & 0.6 & 0.1 & 0.1 \\
    \noalign{\hrule height 2pt} 
  \end{tabular}%
  }
\label{tab:optimization_profile_table}
\end{table}

\noindent
{\bf Baseline Comparisons.}
The primary baseline metaheuristic for comparison is SA. 
To ensure a strong baseline, we conduct an extensive sweep of SA settings and report the best observed result for each workload-profile pair. Each configuration is evaluated by runtime and best observed system cost within the allotted budget. The setting value are as defined in Section~\ref{sec:pareto}. All experiments use OpenAI's GPT-5.3-codex~\cite{openai2026gpt53codex} via the Codex CLI~\cite{codex} with number of parallel running field agents ($N$) set to 100; the model can be easily replaced through a single input parameter without modifying the framework.

%% file: sec/8-results.tex
\begin{figure}[t]
  \centering
  {\includegraphics[width=0.24\textwidth]{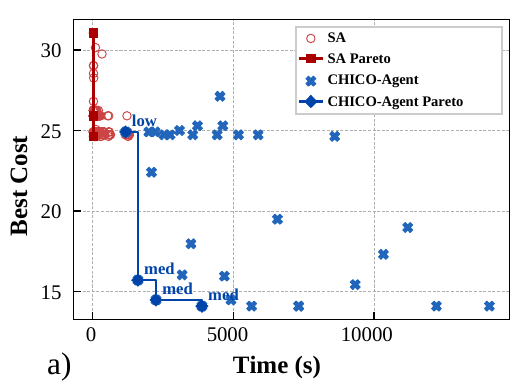}}\hfill
  {\includegraphics[width=0.24\textwidth]{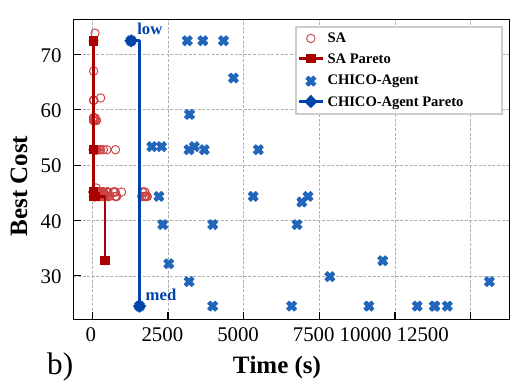}}\hfill
  {\includegraphics[width=0.24\textwidth]{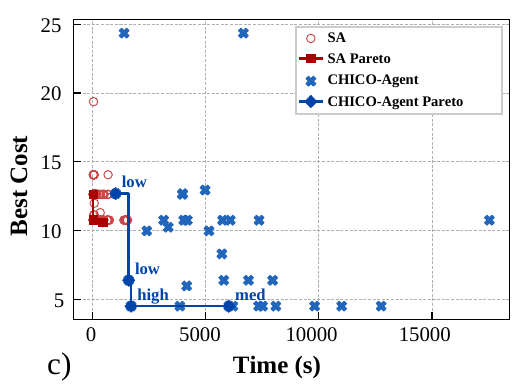}}\hfill
  {\includegraphics[width=0.24\textwidth]{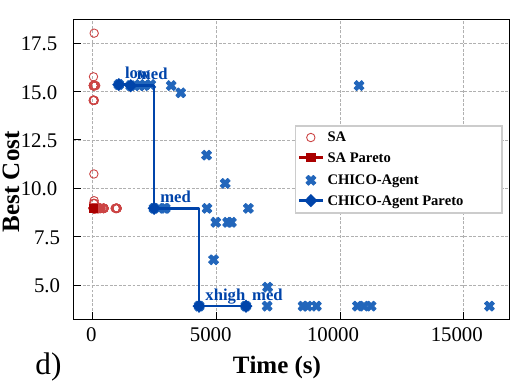}}
  \vspace{-3mm}
  \caption{Runtime vs. cost Pareto frontiers of WL-6 for (a) automotive, (b) balance, (c) mobile, and (d) wearables optimization profiles. All other workloads have similar trends, please refer to our GitHub repository~\cite{CHICO-Agent}.}
  \vspace{-6mm}
  \label{fig:combined_six_wide}
\end{figure}

\section{Results and Evaluation}
\noindent
We compare CHICO-Agent against SA in terms of runtime and generated system cost. Specifically, we plot the tradeoff between wall-clock runtime and converged system cost, and comparing their Pareto-optimal fronts. In addition, we analyze the generated system configurations, i.e., the architecture $x$ defined in Eq.~\eqref{eq:opt_problem}, to understand the design choices made by each method.

\vspace{-1mm}
\subsection{Pareto Frontier Analysis} \label{sec:pareto}
\noindent
To construct the Pareto frontiers, we sweep the hyperparameters of both SA~\cite{carbon-path} and CHICO-Agent, and record the cost and runtime. For SA, we perform a grid search over initial temperature $T_0 \in [4000,,10000]$ (step 500) and cooling rate $r \in [0.70,,0.99]$ (steps of 0.01), yielding 390 configurations per (workload, profile) pair. For CHICO-Agent, we sweep reasoning effort (\texttt{low}, to \texttt{xhigh}) and iteration count $i \in [5,40]$ (steps of 5), with each combination producing one (cost, runtime) point. A configuration \emph{dominates} another if it achieves both lower cost and lower runtime. The Pareto frontier for each method consists of the set of \emph{non-dominated} configurations. Fig.~\ref{fig:combined_six_wide} shows the Pareto frontiers for SA and CHICO-Agent on WL-6 across all four application profiles\footnote{Due to space constraints, we show WL-6 results here; results for all workloads are available in our GitHub repository~\cite{CHICO-Agent} and show similar trends}.

\begin{figure*}
    \centering
    \includegraphics[width=\textwidth]{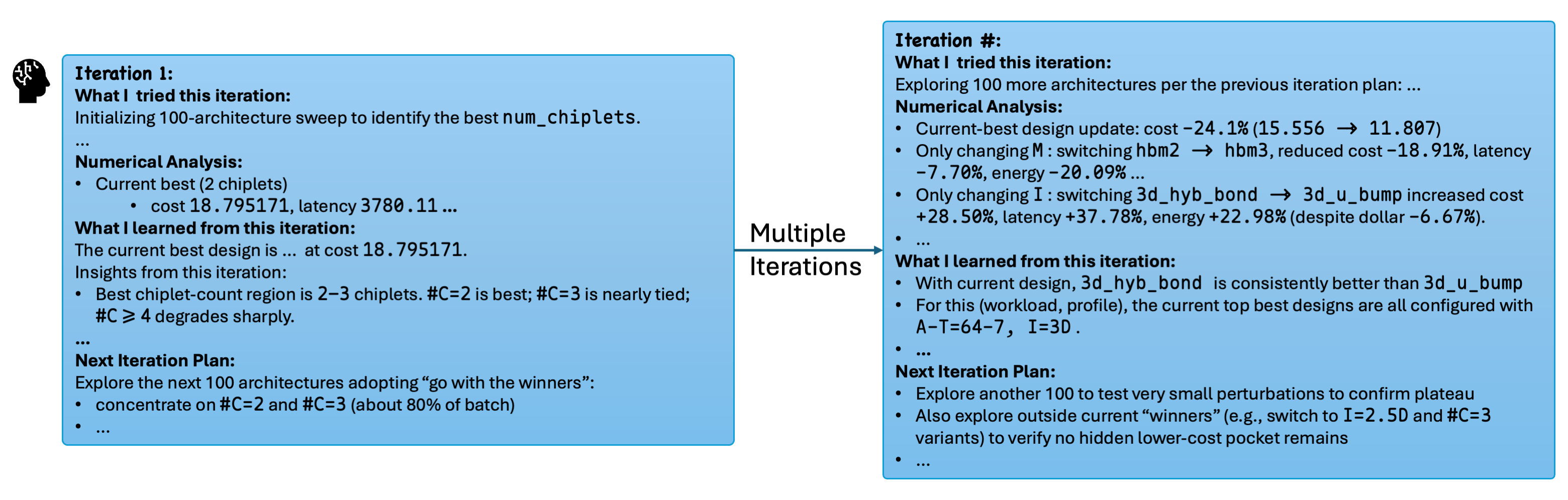}
    \vspace{-8mm}
    \caption{Demonstration of CHICO-Agent evaluation and reasoning outputs, which appended to \texttt{KNOWHOW.md} across iterations.}
    \vspace{-5mm}
    \label{fig:reasoning-trace}
\end{figure*}

Across all profiles, CHICO-Agent consistently identifies configurations with lower cost than the best SA solution in~\cite{carbon-path}. The improvement is particularly pronounced in the automotive and wearables profiles, where CHICO-Agent explores regions of the design space not reached by SA. However, these gains come at the expense of increased runtime. Each CHICO-Agent iteration combines model inference and structured reasoning, which makes it slower than the lightweight perturbation-based updates in SA. Despite this, CHICO-Agent requires significantly less hyperparameter tuning. SA depends on multiple tightly coupled parameters, initial temperature, final temperature, cooling rate, and number of moves per temperature, whose interaction is problem-specific and necessitates an extensive grid search (390 configurations in our evaluation). In contrast, CHICO-Agent exposes only reasoning effort and iteration count, both of which are intuitive and exhibit fewer combinatorial interactions. Furthermore, CHICO-Agent provides textual reasoning , offering an audit trail that explains \emph{why} specific architectural decisions are made—an advantage not present in traditional optimization methods. Examples of reasoning traces are shown in Fig.~\ref{fig:reasoning-trace}. The trace illustrates CHICO-Agent's iterative design space exploration process, where each iteration follows a structured reasoning loop. All such information is continuously appended to \texttt{KNOWHOW.md}, helping human designers to draw insights and understand the evolving design landscape alongside the agent.

Table~\ref{tab:best_cost_summary} summarizes the best cost achieved by each method across all six workloads and four application profiles. CHICO-Agent outperforms SA in 20 out of 24 (workload, profile) pairs, often by a significant margin. SA achieves better results in only four cases. 
Two of the four SA-favorable cases occur under the balance profile (WL-1 and WL-4). After interacting with the LLM and interpreting the reasoning summary, we discovered  that equal weights reduce the directional signal available to CHICO-Agent’s heuristic reasoning. SA optimizes the same scalar objective through stochastic perturbations.


\input{sec/best_cost_achieved_comparison.tex}

\input{sec/arch_compare_tables.tex}

\vspace{-1mm}
\subsection{Architectural Comparison}
\noindent
Table~\ref{tab:arch_compare_all} compares the best-found architectures of SA and CHICO-Agent across all six workloads and four application profiles.
Each row corresponds to one (workload, profile) pair; the application column abbreviates the four cost profiles as Automotive (Auto.), Balance (Bal.), Mobile (Mob.), and Wearables (Wear.).
For each method the table reports chiplet count(\textbf{\#C}), chiplet specification (\textit{A\nobreakdash-T\nobreakdash-S}, where ``x\textit{n}'' denotes \textit{n} identical replicas), workload mapping (\textit{O\nobreakdash-D\nobreakdash-K}), packaging (\textit{I\nobreakdash-P\nobreakdash-M}), and die-to-die protocol (\textbf{Proto}; ``---'' indicates no D2D link is needed, ``/'' indicates all presented D2D links are used), all using the notation defined in Sec.~\ref{sec:background} with NA indicates not applicable. 
Several differences in design strategy emerge from this comparison.

\noindent
\textbf{Dataflow and split-K.}
SA~\cite{carbon-path} selects OS dataflow in 20 of 24 cases, whereas CHICO-Agent overwhelmingly prefers IS or WS, choosing OS only once (WL\nobreakdash-2, mobile). Closely related, CHICO-Agent enables split-K in most cases. This pairing of IS dataflow with split-K partitioning suggests the agent has learned that distributing the reduction dimension across chiplets improves utilization for the workloads tested, a cross-layer insight that purely stochastic search is unlikely to discover systematically.

\noindent
\textbf{Chiplet sizing.}
SA frequently selects large or different-sized chiplet configurations---e.g., \textit{128\nobreakdash-7\nobreakdash-2048} for WL\nobreakdash-2 automotive, or mixed \textit{128\nobreakdash-7\nobreakdash-1024}$\times$4\,+\,\textit{64\nobreakdash-7\nobreakdash-256}$\times$2 for WL\nobreakdash-3 automotive.
CHICO-Agent, by contrast, converges on smaller identical chiplets, predominantly \textit{64\nobreakdash-7\nobreakdash-256}. This preference reduces area footprint (no white space), which directly lowers cost in the packaging-aware objective.

\noindent
\textbf{Packaging and protocol.}
When CHICO-Agent achieves a lower cost than SA, it sometimes does so with simpler packaging. In four of the six WL-1 to WL-3 balance/mobile cases, CHICO-Agent selects monolithic 2D-NA-DDR5 instead of 3D-HB-HBM3 with UCIe 3D; the exceptions are WL-1 mobile and WL-2 mobile. For WL-4 to WL-6, advanced packaging becomes more common, but integration style and memory choice remain workload-dependent, spanning 2.5D-RDL, 3D-HB, and 2.5D+3D configurations. Design-space convergence. WL-5 shows the strongest cross-method convergence, with both methods frequently selecting homogeneous 64-7-256 chiplets and HBM3-based advanced packaging. WL-6 remains mixed, with agreement in some profiles and different packaging choices in others. In these cases, the remaining gap is often attributable to mapping decisions such as dataflow and split-K.

\noindent
\textbf{Design-space convergence.}
WL\nobreakdash-5 and WL\nobreakdash-6 reveal near-identical hardware choices between the two methods---both settle on homogeneous \textit{64\nobreakdash-7\nobreakdash-256} chiplets with \textit{3D\nobreakdash-HB\nobreakdash-HBM3} and UCIe\,3D---differing only in mapping parameters. This convergence indicates a narrow Pareto-optimal hardware region for these smaller workloads, where the agent's advantage materializes primarily through superior software-level decisions (dataflow and split-K selection).

%% file: sec/best_cost_achieved_comparison.tex
\begin{table}[t]
\centering
\caption{Best cost achieved by CHICO-Agent and SA~\cite{carbon-path} with elapsed time.}
\label{tab:best_cost_summary}
\resizebox{\columnwidth}{!}{%
\setlength{\tabcolsep}{2pt}
\begin{tabular}{
  l c r r
  @{\hskip 5pt\vrule width 0.5pt\hskip 5pt}
  l c r r
}
\toprule
\multirow{2}{*}{\textbf{Profile}} &
\multirow{2}{*}{\textbf{WL}} &
{\textbf{Best Cost}} &
{\textbf{Time (s)}} &
\multirow{2}{*}{\textbf{Profile}} &
\multirow{2}{*}{\textbf{WL}} &
{\textbf{Best Cost}} &
{\textbf{Time (s)}} \\
& & {\textbf{CHICO\,/\,SA}} & {\textbf{CHICO\,/\,SA}}
& & & {\textbf{CHICO\,/\,SA}} & {\textbf{CHICO\,/\,SA}} \\
\midrule
\multirow{6}{*}{Auto.} & \cellcolor{llmwin}1 & \cellcolor{llmwin}\textbf{36.62}\,/\,41.92 & \cellcolor{llmwin}2957\,/\,7719 & \multirow{6}{*}{Mob.} & 1 & 8.51\,/\,\textbf{6.74} & 14028\,/\,10132 \\
 & \cellcolor{llmwin}2 & \cellcolor{llmwin}\textbf{28.00}\,/\,39.83 & \cellcolor{llmwin}5415\,/\,3148 & & \cellcolor{llmwin}2 & \cellcolor{llmwin}\textbf{11.76}\,/\,16.21 & \cellcolor{llmwin}13317\,/\,1093 \\
 & \cellcolor{llmwin}3 & \cellcolor{llmwin}\textbf{72.61}\,/\,130.29 & \cellcolor{llmwin}11405\,/\,1576 & & \cellcolor{llmwin}3 & \cellcolor{llmwin}\textbf{11.38}\,/\,25.07 & \cellcolor{llmwin}16112\,/\,10003 \\
 & \cellcolor{llmwin}4 & \cellcolor{llmwin}\textbf{69.15}\,/\,87.01 & \cellcolor{llmwin}13894\,/\,46 & & \cellcolor{llmwin}4 & \cellcolor{llmwin}\textbf{12.95}\,/\,19.60 & \cellcolor{llmwin}9466\,/\,1845 \\
 & \cellcolor{llmwin}5 & \cellcolor{llmwin}\textbf{37.06}\,/\,39.31 & \cellcolor{llmwin}12771\,/\,224 & & \cellcolor{llmwin}5 & \cellcolor{llmwin}\textbf{5.47}\,/\,8.97 & \cellcolor{llmwin}9305\,/\,1020 \\
 & \cellcolor{llmwin}6 & \cellcolor{llmwin}\textbf{14.10}\,/\,24.64 & \cellcolor{llmwin}3887\,/\,29 & & \cellcolor{llmwin}6 & \cellcolor{llmwin}\textbf{4.50}\,/\,10.57 & \cellcolor{llmwin}6024\,/\,458 \\
\midrule
\multirow{6}{*}{Bal.} & 1 & 66.80\,/\,\textbf{59.24} & 3383\,/\,8796 & \multirow{6}{*}{Wear.} & \cellcolor{llmwin}1 & \cellcolor{llmwin}\textbf{8.25}\,/\,12.07 & \cellcolor{llmwin}9627\,/\,4213 \\
 & \cellcolor{llmwin}2 & \cellcolor{llmwin}\textbf{57.57}\,/\,113.08 & \cellcolor{llmwin}7930\,/\,717 & & 2 & 12.86\,/\,\textbf{9.50} & 8968\,/\,2334 \\
 & \cellcolor{llmwin}3 & \cellcolor{llmwin}\textbf{87.81}\,/\,182.52 & \cellcolor{llmwin}9589\,/\,1302 & & \cellcolor{llmwin}3 & \cellcolor{llmwin}\textbf{22.89}\,/\,28.71 & \cellcolor{llmwin}5506\,/\,696 \\
 & 4 & 109.31\,/\,\textbf{94.52} & 14925\,/\,3735 & & \cellcolor{llmwin}4 & \cellcolor{llmwin}\textbf{12.82}\,/\,19.72 & \cellcolor{llmwin}8329\,/\,567 \\
 & \cellcolor{llmwin}5 & \cellcolor{llmwin}\textbf{53.14}\,/\,62.06 & \cellcolor{llmwin}6576\,/\,882 & & \cellcolor{llmwin}5 & \cellcolor{llmwin}\textbf{5.40}\,/\,8.18 & \cellcolor{llmwin}6252\,/\,289 \\
 & \cellcolor{llmwin}6 & \cellcolor{llmwin}\textbf{24.55}\,/\,32.77 & \cellcolor{llmwin}3973\,/\,415 & & \cellcolor{llmwin}6 & \cellcolor{llmwin}\textbf{3.91}\,/\,8.97 & \cellcolor{llmwin}6188\,/\,35 \\
\bottomrule
\end{tabular}%
}
\vspace{-5mm}
\end{table}

%% file: sec/arch_compare_tables.tex
\begin{table*}[htbp]
\centering
\caption{Architecture comparison of SA vs.\ CHICO-Agent best configurations across all workloads and application profiles.}
\vspace{-1mm}
\label{tab:arch_compare_all}
\resizebox{\textwidth}{!}{%
\setlength{\tabcolsep}{2pt}
\scriptsize
\begin{tabular}{
  !{\vrule width 1.5pt} c l
  !{\vrule width 1.5pt} c c c c c
  !{\vrule width 1.5pt} c c c c c
  !{\vrule width 1.5pt}}
\noalign{\hrule height 1.5pt}
& & \multicolumn{5}{!{\vrule width 1.5pt}c!{\vrule width 1.5pt}}{\textbf{SA Best Configuration~\cite{carbon-path}}}
  & \multicolumn{5}{!{\vrule width 1.5pt}c!{\vrule width 1.5pt}}{\textbf{CHICO-Agent Best Configuration}} \\
\cline{3-7}\cline{8-12}
\textbf{WL} & \textbf{Profile}
  & \textbf{\#C} & \textbf{\textit{A-T-S}} & \textbf{\textit{O-D-K}} & \textbf{\textit{I-P-M}} & \textbf{Proto}
  & \textbf{\#C} & \textbf{\textit{A-T-S}} & \textbf{\textit{O-D-K}} & \textbf{\textit{I-P-M}} & \textbf{Proto} \\
\noalign{\hrule height 1.5pt}
\multirow{4}{*}{1} & Auto. & 1 & \textit{96-7-512} & \textit{1-OS-1} & \textit{2D-NA-DDR5} & --- & 1 & \textit{64-7-512} & \textit{1-IS-1} & \textit{2D-NA-DDR5} & --- \\
 & Bal. & 2 & \textit{96-7-1024}, \textit{64-10-768} & \textit{0-OS-1} & \textit{3D-HB-HBM3} & UC3 & 1 & \textit{64-7-1024} & \textit{0-IS-1} & \textit{2D-NA-DDR5} & --- \\
 & Mob. & 2 & \textit{96-7-512}, \textit{64-7-768} & \textit{0-OS-1} & \textit{3D-HB-HBM3} & UC3 & 5 & \textit{64-7-256}\,x5 & \textit{1-IS-1} & \textit{3D-HB-HBM3} & UC3 \\
 & Wear. & 5 & \textit{64-7-256}\,x5 & \textit{0-OS-0} & \textit{3D-HB-HBM3} & UC3 & 5 & \textit{64-7-256}\,x5 & \textit{0-IS-1} & \textit{3D-HB-HBM3} & UC3 \\
\hline
\multirow{4}{*}{2} & Auto. & 1 & \textit{128-7-2048} & \textit{0-WS-0} & \textit{2D-NA-DDR5} & --- & 1 & \textit{96-7-1024} & \textit{0-IS-1} & \textit{2D-NA-DDR5} & --- \\
 & Bal. & 6 & \textit{64-7-256}\,x5, \textit{96-7-512} & \textit{1-OS-0} & \textit{3D-HB-HBM3} & UC3 & 1 & \textit{96-7-512} & \textit{0-WS-1} & \textit{2D-NA-DDR5} & --- \\
 & Mob. & 5 & \textit{64-7-256}\,x3, \textit{96-7-512}\,x2 & \textit{1-OS-0} & \textit{3D-HB-HBM3} & UC3 & 4 & \textit{96-7-512}, \textit{64-7-256}\,x3 & \textit{1-OS-0} & \textit{3D-HB-HBM3} & UC3 \\
 & Wear. & 2 & \textit{64-7-512}, \textit{64-7-256} & \textit{1-OS-1} & \textit{3D-HB-HBM3} & UC3 & 6 & \textit{64-7-256}\,x6 & \textit{0-IS-1} & \textit{3D-HB-HBM3} & UC3 \\
\hline
\multirow{4}{*}{3} & Auto. & 6 & \textit{128-7-1024}\,x4, \textit{64-7-256}\,x2 & \textit{1-OS-0} & \textit{2.5D+3D-HB/RDL-HBM3} & UC3/S & 6 & \textit{96-7-512}\,x6 & \textit{0-WS-1} & \textit{3D-HB-HBM3} & UC3 \\
 & Bal. & 2 & \textit{64-7-512}, \textit{96-7-1536} & \textit{0-IS-1} & \textit{3D-HB-HBM3} & UC3 & 1 & \textit{128-7-1024} & \textit{0-IS-1} & \textit{2D-NA-DDR5} & --- \\
 & Mob. & 6 & \textit{96-7-512}\,x6 & \textit{0-OS-0} & \textit{3D-HB-HBM3} & UC3 & 1 & \textit{128-7-1024} & \textit{1-IS-1} & \textit{2D-NA-DDR5} & --- \\
 & Wear. & 5 & \textit{96-7-512}\,x5 & \textit{0-OS-0} & \textit{3D-HB-HBM3} & UC3 & 6 & \textit{96-7-512}\,x6 & \textit{0-WS-1} & \textit{3D-HB-HBM3} & UC3 \\
\hline
\multirow{4}{*}{4} & Auto. & 2 & \textit{128-7-1024}\,x2 & \textit{1-OS-0} & \textit{2.5D-RDL-HBM3} & UCS & 2 & \textit{128-7-1024}\,x2 & \textit{1-IS-0} & \textit{2.5D-RDL-DDR5} & UCS \\
 & Bal. & 2 & \textit{96-7-512}, \textit{64-7-256} & \textit{1-IS-0} & \textit{2.5D-EMIB-HBM3} & UCA & 4 & \textit{96-7-256}\,x4 & \textit{1-IS-1} & \textit{3D-HB-HBM2} & UC3 \\
 & Mob. & 5 & \textit{96-7-512}\,x4, \textit{64-7-256} & \textit{1-OS-0} & \textit{3D-HB-HBM3} & UC3 & 6 & \textit{64-7-256}\,x6 & \textit{0-IS-1} & \textit{3D-HB-HBM3} & UC3 \\
 & Wear. & 5 & \textit{64-7-256}\,x5 & \textit{0-OS-0} & \textit{3D-HB-HBM3} & UC3 & 6 & \textit{64-7-256}\,x6 & \textit{0-IS-1} & \textit{3D-HB-HBM3} & UC3 \\
\hline
\multirow{4}{*}{5} & Auto. & 4 & \textit{64-7-256}\,x4 & \textit{0-OS-0} & \textit{2.5D+3D-HB/RDL-HBM3} & UC3/S & 4 & \textit{64-7-256}\,x4 & \textit{1-IS-1} & \textit{3D-HB-HBM3} & UC3 \\
 & Bal. & 5 & \textit{64-7-256}\,x5 & \textit{0-OS-0} & \textit{3D-HB-HBM3} & UC3 & 4 & \textit{64-7-256}\,x4 & \textit{0-IS-1} & \textit{3D-HB-HBM3} & UC3 \\
 & Mob. & 4 & \textit{64-7-256}\,x4 & \textit{0-OS-0} & \textit{3D-HB-HBM3} & UC3 & 4 & \textit{64-7-256}\,x4 & \textit{1-IS-1} & \textit{3D-HB-HBM3} & UC3 \\
 & Wear. & 4 & \textit{64-7-256}\,x4 & \textit{0-OS-0} & \textit{3D-HB-HBM3} & UC3 & 4 & \textit{64-7-256}\,x4 & \textit{0-IS-1} & \textit{3D-HB-HBM3} & UC3 \\
\hline
\multirow{4}{*}{6} & Auto. & 3 & \textit{64-7-256}\,x3 & \textit{1-OS-0} & \textit{2.5D+3D-HB/RDL-HBM3} & UC3/S & 2 & \textit{64-7-256}\,x2 & \textit{0-WS-0} & \textit{2.5D-RDL-HBM3} & UCS \\
 & Bal. & 2 & \textit{64-7-256}, \textit{64-7-512} & \textit{0-OS-0} & \textit{2.5D-RDL-HBM3} & UCS & 2 & \textit{64-7-256}\,x2 & \textit{0-WS-0} & \textit{3D-HB-HBM3} & UC3 \\
 & Mob. & 2 & \textit{64-7-512}, \textit{64-7-1024} & \textit{1-WS-0} & \textit{3D-\textmu B-HBM3} & UC3 & 2 & \textit{64-7-256}\,x2 & \textit{1-WS-0} & \textit{3D-HB-HBM3} & UC3 \\
 & Wear. & 2 & \textit{64-7-256}\,x2 & \textit{0-OS-0} & \textit{3D-HB-HBM3} & UC3 & 2 & \textit{64-7-256}\,x2 & \textit{0-WS-0} & \textit{3D-HB-HBM3} & UC3 \\
\noalign{\hrule height 1.5pt}
\end{tabular}%
}
\vspace{-4mm}
\end{table*}

%% file: sec/10-conclusion.tex
\vspace{-1mm}
\section{Conclusion and Future Work}
\noindent
We presented CHICO-Agent, an LLM-driven optimization framework that replaces the stochastic perturbation mechanisms of traditional metaheuristics with a reasoning-driven optimization loop for cross-layer DSE of 2.5D and 3D chiplet-based systems. By leveraging a persistent knowledge base, PPAC analytical models, and a hierarchical admin--field multi-agent workflow, CHICO-Agent achieves a lower system cost than a simulated-annealing baseline on GEMM workloads, while using substantially fewer hyperparameters to tune and producing interpretable reasoning traces that link design-parameter changes to PPAC outcomes. The primary limitations are higher per-iteration runtime due to LLM inference overhead and reduced effectiveness under uniformly weighted cost profiles, where the absence of a dominant objective deprives the agent of a clear reasoning anchor. Future work will focus on reducing runtime and integrating metaheuristic optimization with CHICO-Agent.